\documentclass[preprint,aps,showpacs,amsmath,amssymb]{revtex4}
\usepackage{bm,dcolumn}

\newcommand{\mi}{\mathrm{i}}
\newcommand{\NT}{\mathsf{T}}
\newcommand{\NY}{\mathsf{Y}}
\newcommand{\NS}{\mathsf{S}}

\begin{document}

\title{$SU(3)_F$ Meson Mass Formula from Random Phase Approximation}

\author{Weizhen Deng}
\email{dwz@pku.edu.cn}
\author{Xiaolin Chen}
\author{Dahai Lu}
\author{ShiLin Zhu}
\affiliation{%
Department of Physics, Peking University, BEIJING 100871, CHINA}

\date{\today}

\begin{abstract}
We present an $SU(3)_F$ meson mass formula from random phase
approximation (RPA). Both the mesons of ground-state pseudoscalar
octet and ground-state vector octet are described quite well in
this mass formula. We also estimate current and constituent quark
masses from the na\"{\i}ve quark model and the PCAC relation.
\end{abstract}


\maketitle

In quark models, low-lying states of hadron are made up of three
flavor ($u$, $d$, $s$) of quarks and they are classified according to
the $SU_F(3)$ group. Mesons are bound states of quark-antiquark
pair. The ground-states of mesons (pesudoscalars and vectors) form
octets of $SU_F(3)$ group. The well known Gell-Mann--Okubo mass
formula can be obtained from a simple model
Hamiltonian\cite{Gell-Mann,Okubo}
\begin{equation}
\label{HSU3}
H = H_0 + H_1,
\end{equation}
where $H_0$ is $SU_F(3)$ invariant, and $H_1$ is a \emph{minimal}
$SU_F(3)$ breaking interaction which belongs to $(11)$
representation and also conserves isospin $\NT$ and hypercharge $\NY$
\begin{equation}
H_1 = H_1 \left[ (11) \, \NT=0 \, \NY=0 \right].
\end{equation}
From the Wigner-Eckart theorem, hadron masses can be expressed by
reduced matrix elements. For an $SU(3)$ octet, one obtains the mass
formula
\begin{align}
M(\NT \NY) &=  a + \langle (11) \NT\NY
\mid H_1 \left[ (11) \, \NT=0 \, \NY=0 \right] \mid
(11) \NT\NY \rangle \notag \\
&= a + b \NY + c[ \NT(\NT+1) - \frac14 \NY^2 - 1].
\end{align}
Since a particle and its anti-particle must have the same mass due to
the CPT theorem, for a meson octet, $b=0$ and the mass formula is
reduced to
\begin{equation}
\label{vector-mass}
M(\NT \NY) =
a + c[ \NT(\NT+1) - \frac14 \NY^2 - 1].
\end{equation}

The above mass formula fits the vector meson octet very well.
However, due to the chiral symmetry, the mass formula for the
ground-state octet of pseudoscalar meson changes to the following
mass-squared formula\cite{GaLe,DGH}
\begin{equation}
\label{pseudo-mass}
M^2(\NT \NY) = a_2 + c_2[ \NT(\NT+1) - \frac14 \NY^2 - 1].
\end{equation}

In Ref.~\cite{DWZ03}, we proposed an extension to quark potential
models. The chiral symmetry is considered in this extended model with
random phase approximation (RPA).  For meson structure, a
constituent quark potential model is extened to a Hamiltonian of two
coupled channels with an additional $y-$ channel
\begin{equation}
\left[\begin{array}{cc} k_1^0 + k_2^0 + U & V \\ -V & -k_1^0 -k_2^0 - U 
\end{array}\right]
\left[\begin{array}{c} |x\rangle \\ |y\rangle \end{array}\right]
= E
\left[\begin{array}{c} |x\rangle \\ |y\rangle \end{array}\right].
\end{equation}
$V$ is the coupling potential. In our extension, $V$ takes a simple form
\begin{equation}
V = -\frac{4d}{3} \langle \bm{\sigma}_1
\cdot \bm{\sigma}_2 \rangle,
\end{equation}
with a parameter $d$.  $k_1^0 + k_2^0 + U$ is just the effective
hamiltonian of the constituent quark potential model, consisting of
quark and antiquark kinetic energies plus the quark-antiquark
interaction. Under the $SU_F(3)$ symmetry, we should identify it with
eq.~(\ref{HSU3})
\begin{equation}
k_1^0 + k_2^0 + U = H_0 + H_1 \left[ (11) \, \NT=0 \, \NY=0 \right].
\end{equation}
After we classify the mesons into $SU(3)_F$ multiplets
\begin{subequations}
\begin{align}
|x\rangle &= X |\NT,\NY\rangle ,\\
|y\rangle &= Y |\NT,\NY\rangle ,
\end{align}
\end{subequations}
we then obtain
\begin{equation}
\label{EQM-RPA-simp}
\left[\begin{array}{cc} A  & B \\ -B & -A \end{array}\right]
\left[\begin{array}{c} X \\ Y \end{array}\right]
= E
\left[\begin{array}{c} X \\ Y \end{array}\right],
\end{equation}
where
\begin{align}
A(\NT, \NY) &= a_m + c_m[ \NT(\NT+1) - \frac14 \NY^2 - 1], \\
B(\NS) &= -\frac{4d}{3} [2\NS(\NS+1) - 3 ],
\end{align}
$\NS$ is the spin of the meson.

Finally we obtain a unified mass formula
\begin{equation}
\label{modified-mass-formula}
M(\NT \NY \NS) = \sqrt{\left\{a_m + c_m[ \NT(\NT+1) - \frac14 \NY^2 - 1]
\right\}^2 - \left\{\frac{4d}{3} [2\NS(\NS+1) - 3] \right\}^2},
\end{equation}
with three parameters $a_m$, $c_m$ and $d$.  The fit results to meson
ground-states are shown in Table \ref{table-su3-all}.
\begin{table}
\caption{\label{table-su3-all}%
The fitted masses of meson
ground-states, where we take $a_m=895.0$MeV, $c_m=-82.0 $MeV and
$d=200.3$MeV. The experimental data are taken from Ref.~\cite{PDG2000}.
We also list the fitted results from (a) eq.~(\ref{pseudo-mass}) for the 
pseudoscalar octet with $a_2 =(405.3\text{MeV})^2$ and 
$c_2 = -(381.4\text{MeV})^2$; (b) eq.~(\ref{vector-mass}) for the vector 
octet with $a=853.7$MeV and $c=-84.8$MeV. }
\begin{center}
\begin{ruledtabular}
\begin{tabular}{llll}
$m_\pi = 137.4$MeV & $m_\pi^a = 137.3$MeV & $m_{\pi^0}^{\text{exp}} = 135$MeV 
& $m_{\pi^+}^{\text{exp}} = 139.6$MeV \\
$m_K = 483.7$MeV & $m_K^a = 486.8$MeV & $m_{K^+}^{\text{exp}} = 493.7$MeV 
& $m_{K^0}^{\text{exp}} = 497.7$MeV \\
$m_\eta = 559.0$MeV & $m_\eta^a = 556.5$MeV 
& $m_\eta^{\text{exp}}=547.3$MeV & \\
$m_\rho = 767.9$MeV & $m_\rho^b = 768.9$MeV
& $m_\rho^{\text{exp}} = 769.3$MeV & \\
$m_{K^*} = 897.1$MeV & $m_{K^*}^b = 896.2$MeV & 
$m_{K^{*+}}^{\text{exp}} = 891.7$MeV & $m_{K^{*0}}^{\text{exp}} = 896.1$MeV \\
\multicolumn{2}{l}{$(m_\omega+2m_\phi)/3 = 939.8$MeV} 
& $m_\omega^{\text{exp}}=782.6$MeV & $m_\phi^{\text{exp}}=1019.4$MeV \\
\multicolumn{2}{l}{$(m_\omega^b+2m_\phi^b)/3 = 938.6$MeV} &
\multicolumn{2}{l}{$(m_\omega^{\text{exp}}+2m_\phi^{\text{exp}})/3 = 940.5$MeV}
\end{tabular}
\end{ruledtabular}
\end{center}
\end{table}

Both the individual mass formula for the ground-state octet of
pesudoscalar meson eq.~(\ref{pseudo-mass}) and for the vector
meson octet eq.~(\ref{vector-mass}) can be obtained from this unified
mass formula.  For pseudoscalar mesons, $\NS=0$,
\begin{equation}
M_P(\NT \NY) = 
\sqrt{\left\{a_m + c_m[ \NT(\NT+1) - \frac14 \NY^2 - 1]\right\}^2
 -(4d)^2}.
\end{equation}
Since roughly $a_m \approx 4d$, we have
\begin{equation}
M_P^2(\NT \NY) \approx 8c
\left\{a_m + c_m[ \NT(\NT+1) - \frac14 \NY^2 - 1] -(4d) \right\}.
\end{equation}
For vector mesons, $\NS=1$,
\begin{equation}
M_V(\NT \NY) = 
\sqrt{\left\{a_m + c_m[ \NT(\NT+1) - \frac14 \NY^2 - 1]\right\}^2 
- \left(\frac{4d}{3}\right)^2}.
\end{equation}
With $\left(\frac{4d}{3}\right)^2 \ll a_m^2$, thus
\begin{equation}
M_V(\NT \NY) \approx
a_m + c_m[ \NT(\NT+1) - \frac14 \NY^2 - 1].
\end{equation}

Next, We estimate the constituent quark masses from the na\"{\i}ve quark
model.  The vector meson mass formula eq.~(\ref{vector-mass}) can be
understood by assuming that each constituent quark has certain
constituent masse and neglecting the interaction between quark
and antiquark. one has
\begin{align}
m_\rho &= m_u + m_d \approx 2 m_u, \\
m_{K^*} &= m_u + m_s, \\
m_\omega &= 2m_u, \\
m_\phi &= 2m_s.
\end{align}
$m_u$, $m_d$ and $m_s$ are the constituent quark masses of $u$, $d$
and $s$ quarks respectively.  One can estimate from the fitting parameters
\begin{subequations}
\begin{align}
m_u &= \frac{a_m+c_m}{2} =406.5\text{MeV}, \\
m_s &= \frac{a_m}{2} - c_m =529.5\text{MeV}.
\end{align}
\end{subequations}

We can also estimate the current quark masses from the PCAC relation
\begin{equation}
\partial_\mu \left[ \bar\psi_i(x) \gamma^\mu \gamma_5 \psi_j(x) \right]
 = (m_i^0 + m_j^0) \bar\psi_i(x) \mi \gamma_5 \psi_j(x),
\end{equation}
where $m^0$ are the current quark masses.  Let us calculate its matrix
element between $\langle 0 \mid$ and the pesudoscalar meson state
$\mid P\rangle$ which consists of an $i$- quark and a $j$-
antiquark. For the meson at rest, one obtains
\begin{equation}
-\mi m_P \langle 0 \mid \bar\psi_i(0) \gamma^0 \gamma_5 \psi_j(0)
\mid P \rangle =
(m_u+m_d) \langle 0 \mid \bar\psi_i(0) \mi \gamma_5 \psi_j(0) \mid P \rangle.
\end{equation}
In the non-relativistic limit,
\begin{subequations}
\begin{equation}
\langle 0 \mid
\bar\psi(0)_i \gamma^0 \gamma_5 \psi(0)_j \mid P \rangle
= \sqrt2 \left(
\langle \bm{r}=0, S=0,M_S=0 \mid x \rangle
+ \langle \bm{r}=0, S=0,M_S=0 \mid y \rangle \right),
\end{equation}
and
\begin{equation}
\langle 0 \mid
\bar\psi(0)_i \gamma_5 \psi(0)_j \mid P \rangle
= -\sqrt2 \left(
\langle \bm{r}=0, S=0,M_S=0 \mid x \rangle
- \langle \bm{r}=0, S=0,M_S=0 \mid y \rangle \right),
\end{equation}
\end{subequations}
where RHS of equations are connected to the meson spatial wavefunctions at
origin $\bm{r}=0$.  The sum of current masses of the 
quark antiquark pair is
\begin{align}
m_i^0 + m_j^0 &=  \frac{X_P + Y_P}{X_P - Y_P}  m_P
\notag \\
&= \sqrt{ \frac{a_m + c_m[ \NT(\NT+1) - \frac14 \NS^2 - 1] -4d}
{a_m + c_m[ \NT(\NT+1) - \frac14 \NS^2 - 1] +4d}} m_P \notag \\
&= a_m + c_m[ \NT(\NT+1) - \frac14 \NS^2 - 1] -4d .
\end{align}
Then we can estimate the current quark masses
\begin{subequations}
\begin{align}
m_u^0 &= \frac{a_m-4d+c_m}{2} = m_u - 2d = 5.8\text{MeV} \\
m_s^0 &= \frac{a_m-4d}{2} - c_m = m_s - 2d = 128.8\text{MeV}.
\end{align}
\end{subequations}
In current calculation, we do not consider the iso-spin breaking effect from
difference between $m_u^0$ and $m_d^0$ and from the electro-magnetic
interaction.

\section*{Acknowledgements}
This project was supported by the National Natural Science
Foundation of China under Grant 10375003, Ministry of Education of China, 
FANEDD and SRF for ROCS, SEM.


\end{document}